\documentclass[aps,prd,amsmath,onecolumn,notitlepage,showpacs,superscriptaddress,nofootinbib,usenatbib,11pt]{revtex4-1}
\def \be {\begin{equation}} 
\def \ee {\end{equation}} 
\def \bea {\begin{eqnarray}} 
\def \eea {\end{eqnarray}} 

\usepackage{graphicx}
\usepackage{dcolumn}
\usepackage{bm}
\usepackage{epsfig} 
\usepackage{amsfonts}
\usepackage{amsmath}
\usepackage{amssymb}
\usepackage[usenames]{color}
\usepackage[dvipsnames]{xcolor}
\usepackage[unicode, colorlinks=true, linkcolor=linkcolor, citecolor=linkcolor, filecolor=linkcolor,urlcolor=linkcolor, pdfusetitle]{hyperref}
\usepackage{listings}

\hypersetup{colorlinks,citecolor=blue,linkcolor=blue,urlcolor=blue}
\hypersetup{final=true}

\definecolor{lgreen}{HTML}{079d90}

\begin{document}

\title{A null test of the Cosmological Principle with BAO measurements}

\author{Carlos Bengaly}
\email{carlosbengaly@on.br}
\affiliation{Observat\'orio Nacional, 20921-400, Rio de Janeiro - RJ, Brazil}

\date{\today}

\begin{abstract}
The assumption of a Friedmann-Lemaître-Roberton-Walker (FLRW) Universe is one of the fundamental pillars of modern Cosmology. It is crucial to confront it against cosmological observations in order to confirm (or rule out) the validity of the $\Lambda$CDM paradigm. We perform a consistency test of the FLRW metric using measurements of both radial and transversal modes of Baryonic Acoustic Oscillations in a model-independent fashion using Gaussian Processes. The impact of different prior assumptions for the sound horizon scale and the Hubble Constant is also explored. We find a mild deviation of the FLRW hypothesis ($\gtrsim 3\sigma$ confidence level) at lower redshift ranges ($0.1<z<0.3$) under the assumption the SH0ES $H_0$ prior, which does not occur for the $H_0$ prior by Planck. We also find that different reconstruction kernels reduce the statistical significance of such a deviation. At higher reshift ranges, the null test are able to confirm the FLRW assumption within $3\sigma$ confidence level regardless of the priors.  
\end{abstract}

\maketitle


\section{Introduction}\label{sec:intro}

The flat $\Lambda$CDM model has been established as the standard model of Cosmology (SCM) for over two decades now. It describes an Universe dominated by cold dark matter and the Cosmological Constant $\Lambda$, where the former is needed to explain cosmic structure formation and galaxy dynamics, and the latter is the best candidate to explain the so-called dark energy, i.e., an exotic fluid with negative equation of state that is responsible for the late-time cosmic accelerated expansion. The $\Lambda$CDM scenario is based upon two fundamental assumptions: General Relativity as the underlying theory of gravity, in addition to statistical large-scale homogeneity and isotropy at large scales. The latter corresponds to the Cosmological Principle (CP), which allows us to describe cosmic distances and clocks by the Friedmann-Lemaître-Robertson-Walker (FLRW) metric. Although such a model has been tremendously successful to explain cosmological observations from the Cosmic Microwave Background (CMB)~\cite{planck21}, Type Ia Supernovae (SNe) luminosity distances~\cite{pantheon18}, as well as the clustering and weak lensing of the large-scale structure~\cite{eboss21, kids21, des21a, des21b}, it suffers from fine-tuning problems that could not be solved as yet. Moreover, the so-called SH0ES tension between Hubble Constant measurements from SNe and CMB observations may hint at a potential departure of the SCM~\cite{efstathiou21, riess21} - an interested reader is referred to~\cite{divalentino21,perivolaropoulos21,shah21} for an extensive review on the SH0ES tension and other cosmological puzzles.  

As previously said, the CP constitutes one of the most fundamental assumptions of the SCM. Still, it has been seldom confronted with cosmological observations to scrutinise its empirical validity. Testing the CP is a crucial task to underpin - or to rule out - the $\Lambda$CDM paradigm, as well as most of its competitive alternatives. However, a difficulty arises: only the assumption of cosmic isotropy can be directly tested with observations, but homogeneity cannot, since we perform these observations down the past light-cone instead of the time-constant hypersurfaces - which encompass all objects at a single redshift~\cite{clarkson10, maartens11, clarkson12}\footnote{Some tests can circumvent this issue, as proposed in~\cite{heavens11, jimenez19}, but will be pursued in future works.}. 

Nevertheless, there are tests that have been designed to rule out the CP by means of consistency relations between cosmic distances and chronometers. One of these tests consists on comparing measurements of the radial and transverse modes of baryonic acoustic oscillations (BAO) across a given redshift range, as proposed by~\cite{maartens11}. If we find strong inconsistencies between these measurements, they will hint at a violation of the CP - and thus require a profound reformulation of the standard model. In light of recent works that pointed out a potential FLRW breakdown~\cite{Bengaly:2017slg,Colin:2019opb,Migkas:2020fza,Secrest:2020has,cai21,Krishnan:2021dyb,Krishnan:2021jmh,Singal:2021crs,Singal:2021kuu}, our goal is to verify the feasibility of these results using different data and methods. 

This paper is organised as follows: Section 2 is dedicated to describe the data and methodology herein deployed, section 3 presents the results, while section 4 contains the concluding remarks. 

\section{Data and methodology}

We follow the approach proposed by~\cite{arjona21a} (see~\cite{lhuillier17} for a similar test), where the radial and transverse BAO modes are respectively given by
\begin{eqnarray}\label{eq:radial_bao}
L_\parallel = \frac{r_{\rm s}}{D_{\rm C}(z)} \,;
\end{eqnarray}

\begin{eqnarray}\label{eq:transv_bao}
L_\perp = \frac{r_{\rm s}}{(1+z)D_{\rm A}(z)} \,,
\end{eqnarray}
where $r_{\rm s}$ denotes the sound horizon scale at the baryon drag epoch, and $D_{\rm C}(z)$ and $D_{\rm A}(z)$ correspond to the comoving radial and angular diameter distances, respectively, which read~\cite{hogg99}
\begin{eqnarray}\label{eq:Dz}
D_{\rm C}(z) = c\int_0^z \frac{dz'}{H(z)} \,, \quad D_{\rm A}(z) = \frac{D_{\rm C}(z)}{(1+z)}   \,; 
\end{eqnarray}

\begin{eqnarray}\label{eq:Hz}
H(z) = H_0\sqrt{\Omega_{\rm m}(1+z)^3 + \Omega_\Lambda}
\end{eqnarray}
for a FLRW Universe assuming a flat $\Lambda$CDM framework. We neglect radiation and neutrinos in Eq.~\eqref{eq:Hz} as their contribution is very small in the redshift ranges of our analysis.

Hence we can define the null test according to~\cite{maartens11, arjona21a}
\begin{eqnarray}\label{eq:zetaz}
\zeta(z) = 1 - \frac{L_\parallel}{L_\perp} = 1 - \frac{(1+z)D_{\rm A}(z)}{D_{\rm C}(z)}\,,
\end{eqnarray}
so that
\begin{equation}
\zeta(z) \neq 0 \;\; \mbox{implies FLRW ruled out.}    
\end{equation}

This test has an advantage over the $\mathcal{O}_{\rm k}$ test proposed by~\cite{clarkson08} because it does not require the computation of cosmic distances derivatives, which significantly degrades the errors in the analysis. Similar tests were proposed ~\cite{sahni08, february13, rasanen15}, but using different approaches. 

Our data sets consist on 18 $H(z)$ measurements of the radial BAO mode, as compiled at table I of ~\cite{magana18}\footnote{Note that we only use the data points labelled as 'clustering' in our main analysis, and we neglected the covariance between some of these measurements.}, and 15 transverse BAO mode measurements taken from Table I of~\cite{nunes20}. These data points were obtained from SDSS-III luminous red galaxies~\cite{carvalho16, alcaniz17, carvalho20} and quasars~\cite{decarvalho18}, in addition to SDSS-IV blue galaxies~\cite{decarvalho21}. Conversely from the radial BAO, these measurements do not require the assumption of a fiducial model to extract the BAO signal, so we can avoid {\it a priori} assumptions of the underlying cosmology before proceeding with the analysis. Nonetheless, the assumption of a sound horizon scale is required to convert them into angular distance measurements. We assume the $r_{\rm s}$ constraints reported by~\cite{carvalho20} and by~\cite{verde17} (henceforth C20 and VBHJ17, respectively) in our analysis 
\begin{equation}\label{eq:rs}
r^{\rm C20}_{\rm s} = 107.4 \pm 1.7 \, \mathrm{Mpc \; h}^{-1} \,, \\
r^{\rm VBHJ17}_{\rm s} = 101.0 \pm 2.3 \, \mathrm{Mpc \; h}^{-1} \,;\\
\end{equation}
in addition to two different priors on $H_0$, namely from Planck~\cite{planck21} (hereafter P18) and SH0ES latest measurements~\cite{riess21} (hereafter R21), which respectively correspond to
\begin{eqnarray}\label{eq:H0_priors}
H^{\rm P18}_0 = 67.50 \pm 0.50 \; \mathrm{km \; s}^{-1} \; 
\mathrm{Mpc}^{-1} \,; \\
H^{\rm R21}_0 = 73.04 \pm 1.04\; \mathrm{km \; s}^{-1} \; 
\mathrm{Mpc}^{-1} \,.
\end{eqnarray}
Our choice for $r^{\rm C20}_{\rm s}$ is justified by the fact it was obtained directly from most of the transverse BAO measurements adopted this analysis, although with the limitation of the flat $\Lambda$CDM assumption~\cite{carvalho21}. We also test whether the $H_0$ prior can impact our results in light of the hints at a FLRW departure discussed in~\cite{Krishnan:2021dyb,Krishnan:2021jmh}, for instance, as a potential explanation for the SH0ES tension. 

We carry out a non-parametric reconstruction of both $H(z)$ and $D_{\rm A}(z)$ data adopting the Gaussian Processes (GP) method as in the GaPP package (Gaussian Processes in Python)~\citep{seikel12}- see also~\citep{shafieloo12}). Basically, a Gaussian Process consists on the extension of the idea behind Gaussian distributions to general functions, where we assume a covariance function to connect the available data points to other points in space in order to perform a regression analysis. A typical choice for this covariance function corresponds to the squared exponential kernel, as given by
\begin{equation}\label{eq:kernel_sqexp}
k(x,x') = \sigma^2 \exp{\left(-\frac{(x-x')^2}{2l^2}\right)} \,,
\end{equation}
where $l$ and $\sigma$ are the hyperperameters of the GP reconstruction, while $x$ and $x'$ denotes two different points in space. $l$ can be interpreted as its variance, which provides the width of the reconstructed curve, and $\sigma$ quantifies how much it deviates from the average. We assume the kernel shown in Eq.~\eqref{eq:kernel_sqexp} in 1000 reconstruction bins along $0.1 < z < 2.4$ - which corresponds to the whole redshift range covered by the data - as our default choice for the GP reconstructions. We also opt not to optimise the GP hyperparameters in order to avoid possible overfitting. Note that similar methods and applications were pursued in~\cite{shafieloo10, yahya13, sapone14, busti14, costa15, gonzalez16, joudaki17, yu18, marra18, gomezvalent18a, haridasu18, gomezvalent18b, keeley20, bengaly20a, bengaly20b, arjona20, singirikonda20, bengaly21, benisty21, mukherjee20a, mukherjee20b, briffa20, vonMarttens21, colgain21, perenon21, escamilla-rivera21, bernardo21a, aizpuru21, bernardo21b, bernardo21c, dialektopoulos21}, but with different goals and data-sets. This prescription provides a model-independent test, i.e., without any assumption about a cosmological model {\it a priori}, since we are not fitting parametric functions across the data, but rather reconstructing the best curve that passes through the data points using the GP regression - thus our only model-dependency occurs at the GP kernel choice.

Moreover, we numerically integrate the reconstructed $H(z)$ curve in order to obtain $D_{\rm C}$ using trapezoid rule following~\cite{holanda13}, so that
\begin{eqnarray}\label{eq:dc_trap}
D_{\rm C} = c \int_0^z{dz' \over H(z')} \approx \frac{c}{2}\sum_{i=1}^{N} (z_{i+1}-z_i)\left[ {1\over H(z_{i+1})}+{1\over H(z_i)} \right] \,.
\end{eqnarray}
Since the error on $z$ measurements is negligible, we only take the uncertainty on the values of $H(z)$ into account. The uncertainty associated to the $i^{\rm th}$ bin is given by
\begin{equation}
\sigma_i={c\over 2}(z_{i+1}-z_i)\left({\sigma_{H_{i+1}}^2\over H_{i+1}^4} + {\sigma_{H_{{i}}}^2\over H_{i}^4}\right)^{1/2} \,.
\end{equation}
Since this integration is carried out along the evenly spaced-out 1000 bins at $z \in [0.1,2.4]$, rather than the 18 individual measurements across an uneven redshift range, we can circumvent possible numerical limitations in this procedure. 
\section{Results}

\begin{figure*}[!h]
\includegraphics[width=0.48\textwidth,height=7.2cm]{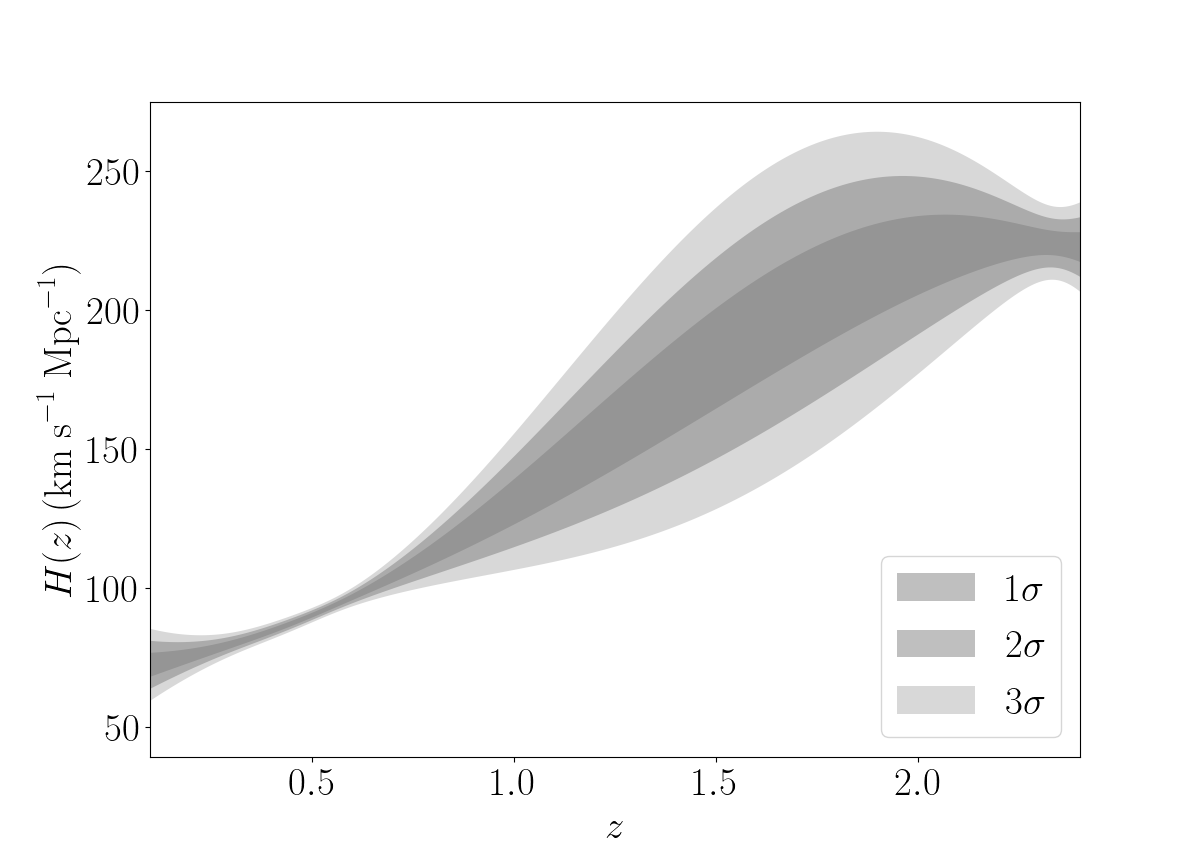}
\includegraphics[width=0.48\textwidth,height=7.2cm]{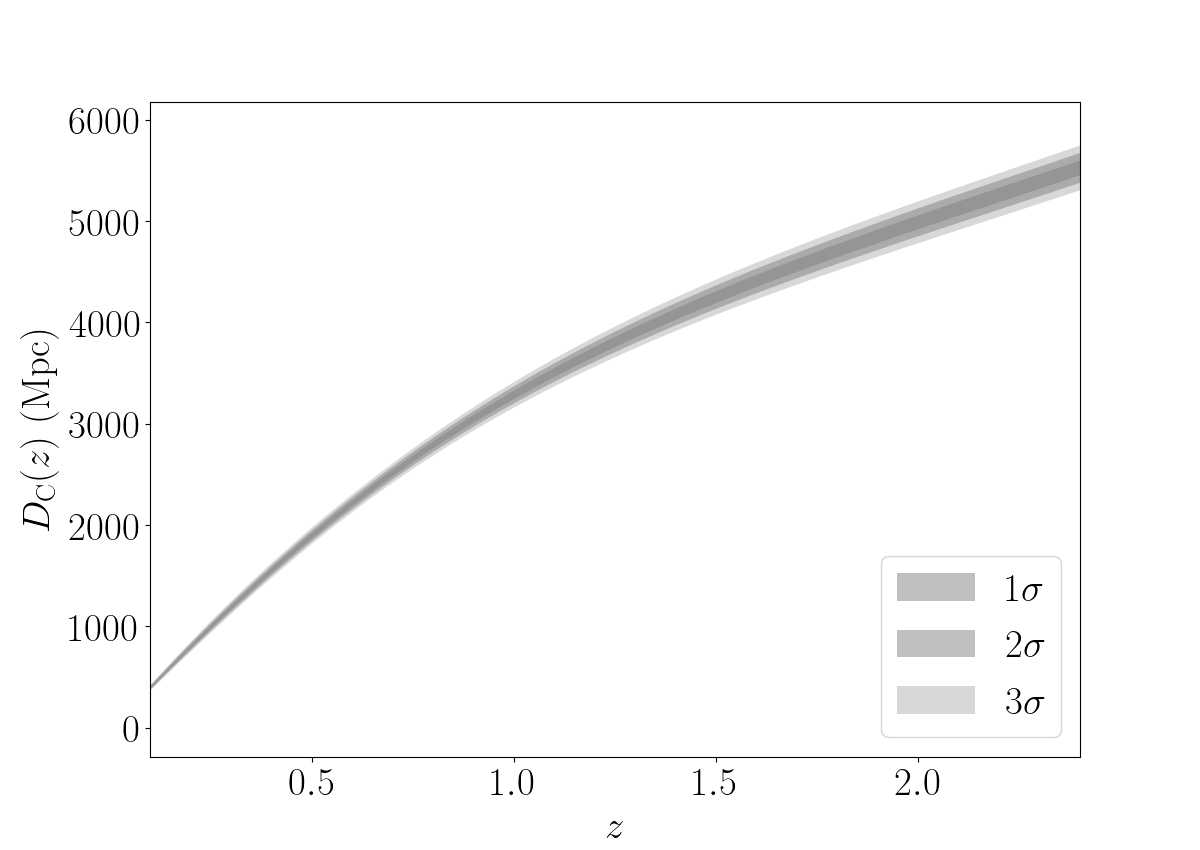}
\caption{The reconstructed $H(z)$ (left panel) and $D_{\rm C}(z)$ (right panel) curves from radial BAO mode measurements, respectively. Different tones of grey denote different confidence levels, ranging from $1$ to $3\sigma$.}
\label{fig:rec_radial_bao}
\end{figure*}

\begin{figure*}[!h]
\includegraphics[width=0.48\textwidth,height=7.2cm]{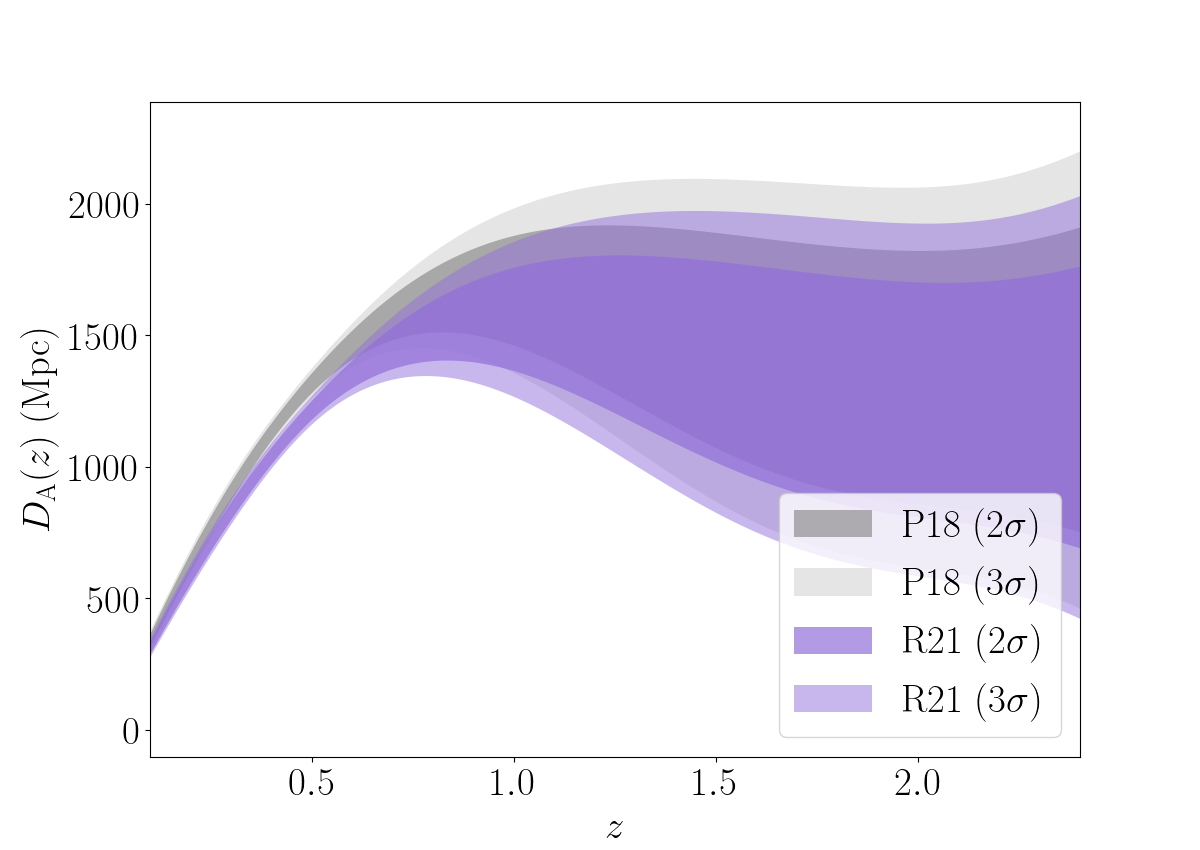}
\includegraphics[width=0.48\textwidth,height=7.2cm]{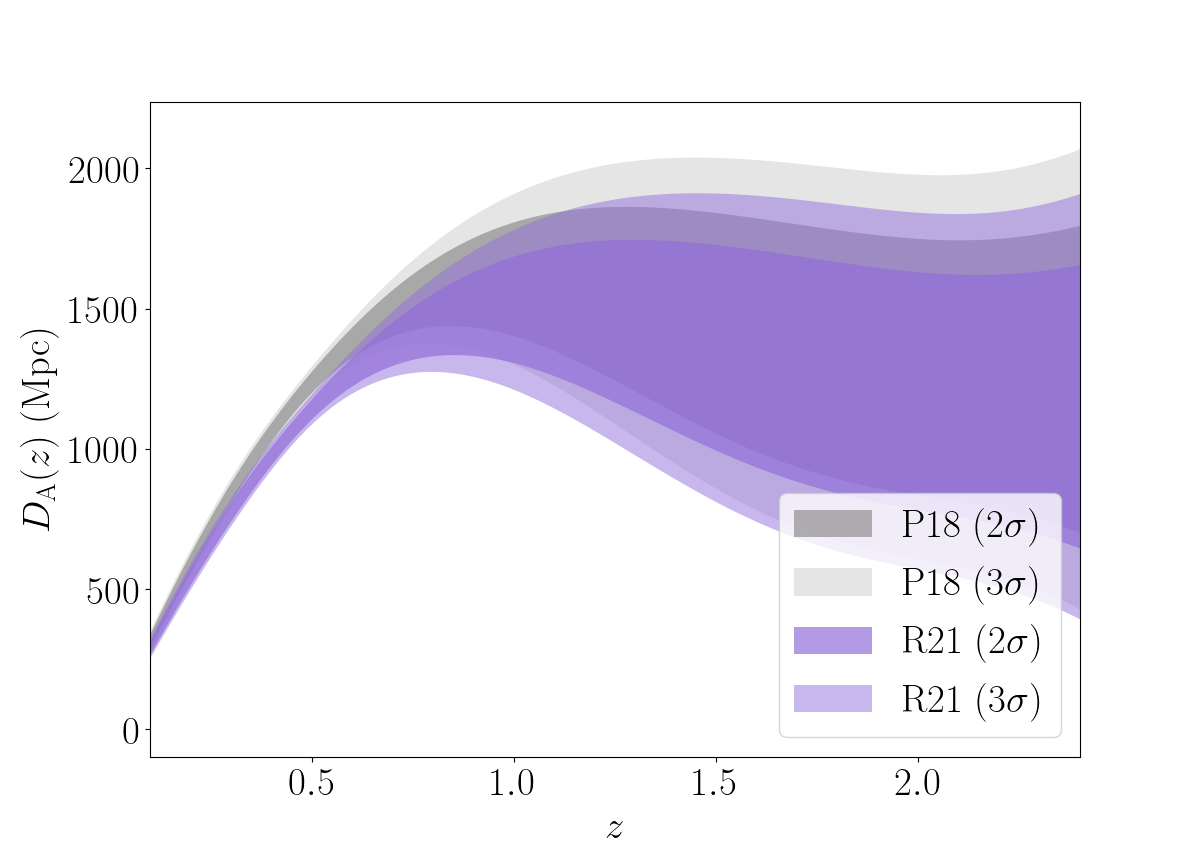}
\caption{The reconstructed $D_{\rm A}(z)$ curves from transverse BAO mode measurements. The left panel was obtained under the assumption of the C20 combined with the P18 (grey shade) and R21 (purple shade) priors. Different tones of purple and grey denote different confidence levels - $2$ and $3\sigma$ in this case. The right panel presents the $D_{\rm A}(z)$ reconstruction assuming the VBJH17 $r_{\rm s}$ prior instead of C20.}
\label{fig:rec_transv_bao}
\end{figure*}

\begin{figure*}[!h]
\includegraphics[width=0.48\textwidth, height=7.2cm]{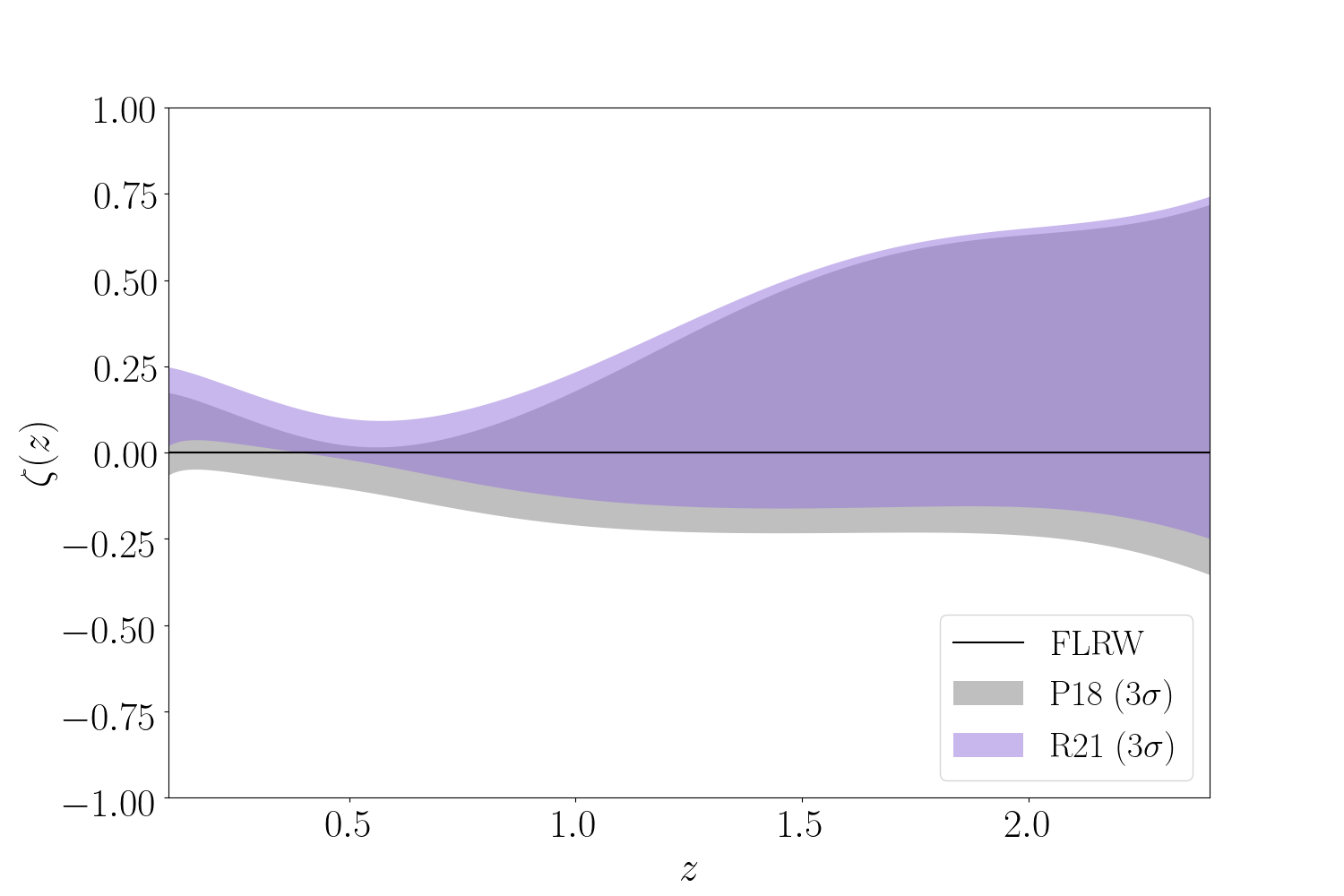}
\includegraphics[width=0.48\textwidth, height=7.2cm]{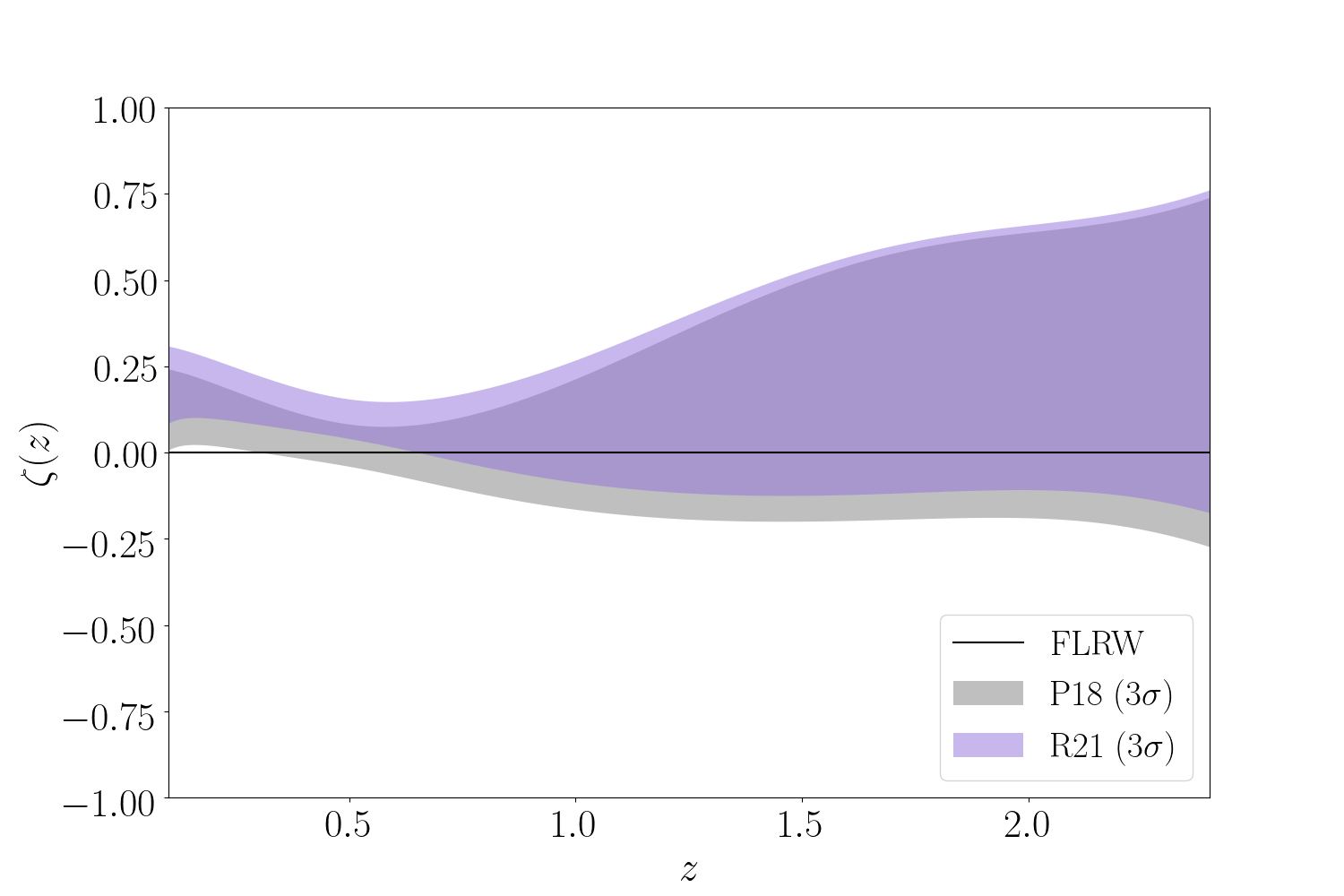}
\caption{{\it Left panel:} Results for the null FLRW test $\zeta(z)$ at a $3\sigma$ CL assuming the C20 $r_s$ measurement, as well as the R21 (purple) and P18 (grey) $H_0$ priors. {\it Right panel:} Same as the previous one, but assuming the VBHJ17 prior instead.}
\label{fig:rec_null_test}
\end{figure*}

\newpage

We show the {\sc GaPP}-reconstructed $H(z)$ (left panel) and $D_{\rm C}(z)$ (right panel) curves from radial BAO data in Fig.~\ref{fig:rec_radial_bao}, at a $1-3\sigma$ confidence level (CL). On the other hand, Fig.~\ref{fig:rec_transv_bao} presents the $D_{\rm A}(z)$ reconstructions obtained from the transverse BAO measurements, where the left panel assumes the C20 $r_{\rm s}$ prior combined with P18 (grey) and R21 (purple) $H_0$ ones, and the right panel assumes the VBHJ17 prior instead. The latter plots are shown at a $2$ and $3\sigma$ CL in order to ease visualisation.

The results of the FLRW null test $\zeta$ are presented in Fig.~\ref{fig:rec_null_test} ($3\sigma$ CL) for C20 (left panel) and VBHJ17 (right panel) priors along with the P18 (grey) and R21 (purple) ones. Our results are compatible with those reported in Fig. 2 of~\cite{arjona21a}, albeit with slightly larger uncertainties as we are using a different reconstruction method, as well as different combination of data-sets and priors on the $r_{\rm s}$ and $H_0$.  
We note a $\approx 3.5\sigma$ deviation from the FLRW assumption around $0.1 < z < 0.3$ when we assume the C20 and R21 priors, which does not occur for the P18 case. On the other hand, this departure occurs for both $H_0$ priors under the assumption of the VBHJ17 $r_{\rm s}$ measurement, which is extended to the $0.1<z<0.8$ redshift range in the VBHJ17-R21 combination. 

\begin{figure*}[!h]
\includegraphics[width=0.48\textwidth, height=7.2cm]{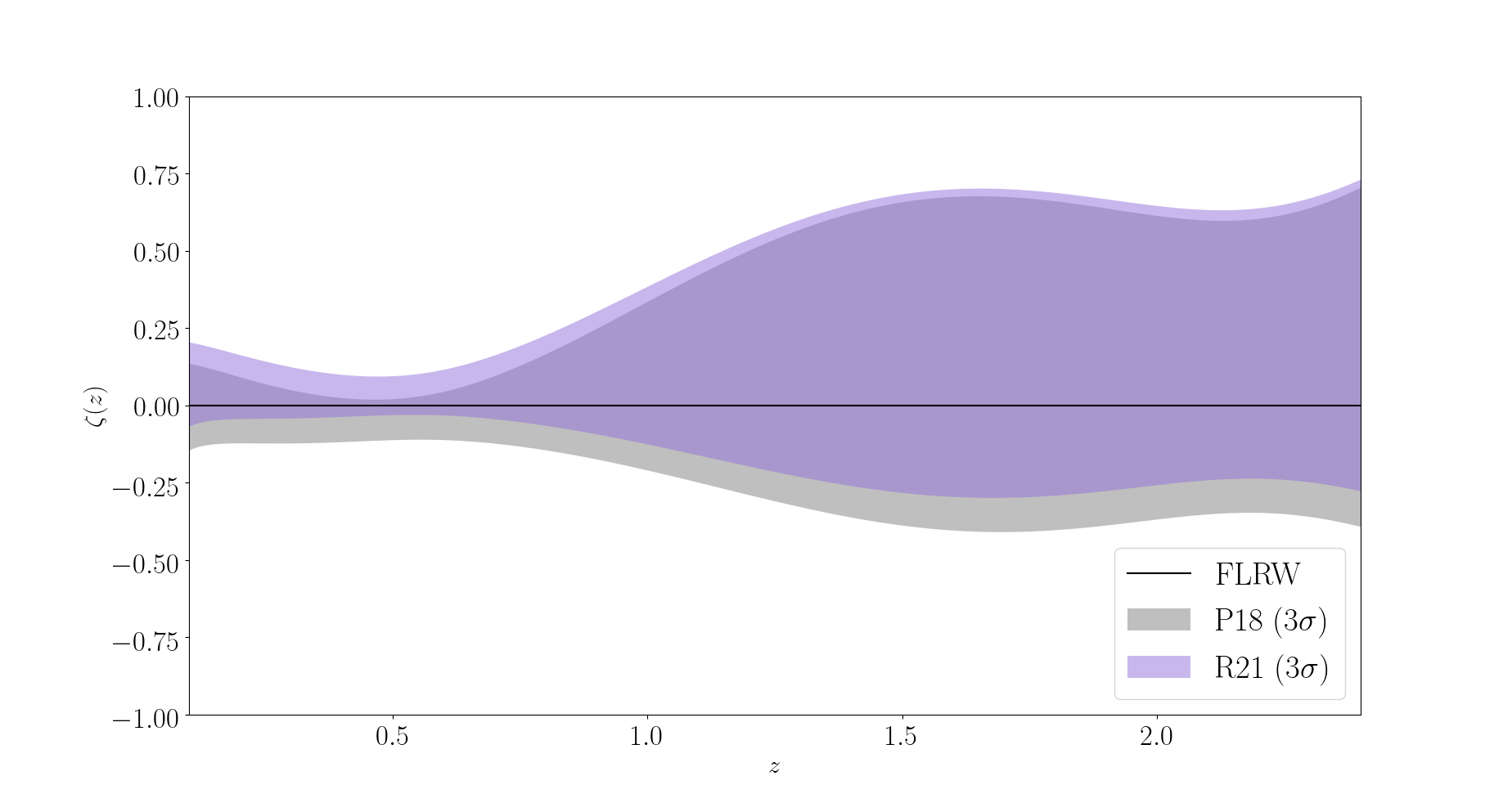}
\includegraphics[width=0.48\textwidth, height=7.2cm]{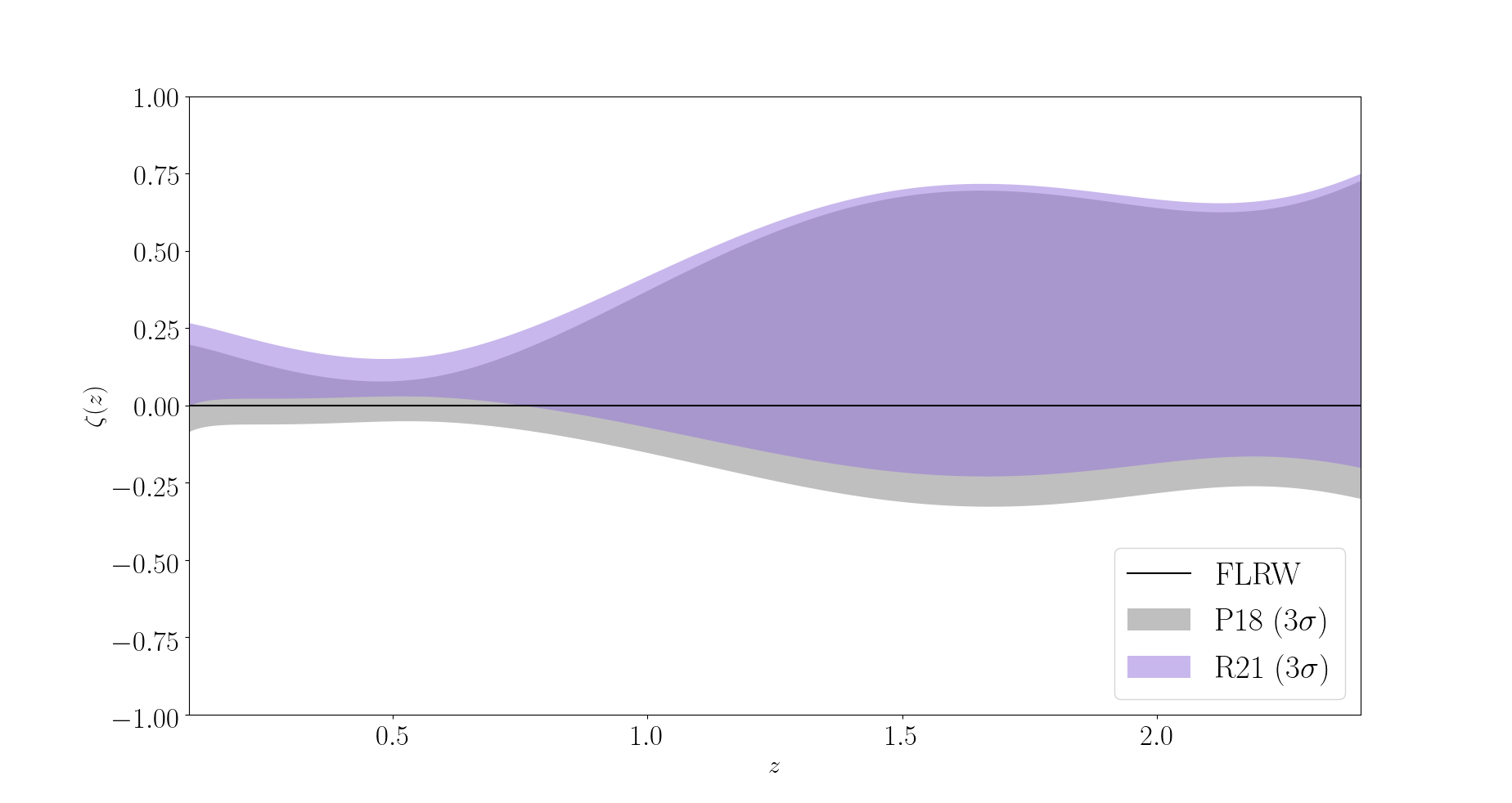}
\caption{{\it Left panel:} Results for the null FLRW test $\zeta(z)$ at a $3\sigma$ CL assuming the C20 $r_s$ measrument, as well as the R21 (purple) and P18 (grey) $H_0$ priors, but assuming the Mat\'ern(9/2) kernel rather than the squared exponential one for the $H(z)$ and $D_A(z)$ reconstructions. {\it Right panel:} Same as the previous one, but assuming the VBHJ17 prior rather than C20.}
\label{fig:rec_null_test_kernel1}
\end{figure*}

\begin{figure*}[!h]
\includegraphics[width=0.48\textwidth, height=7.2cm]{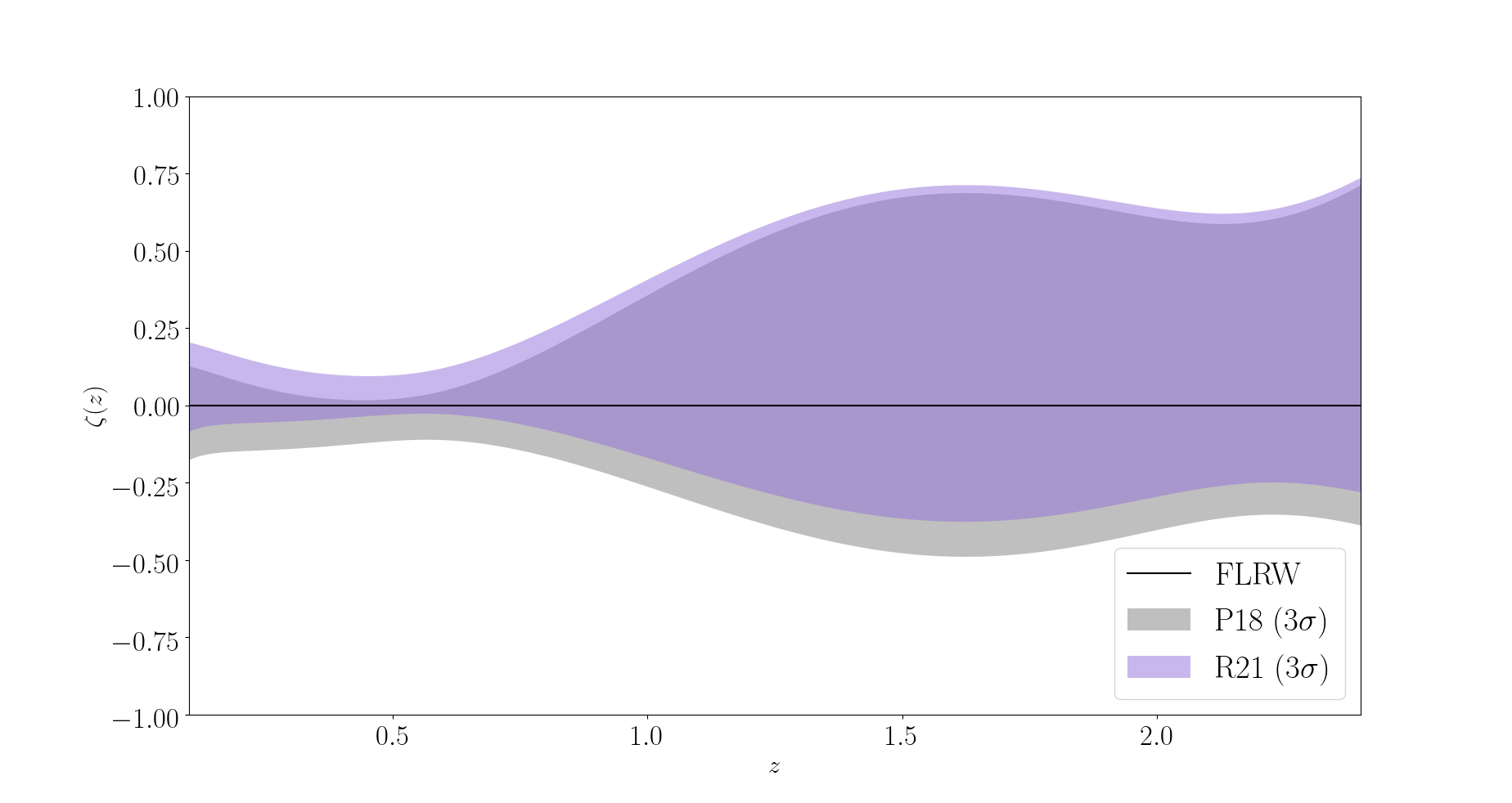}
\includegraphics[width=0.48\textwidth, height=7.2cm]{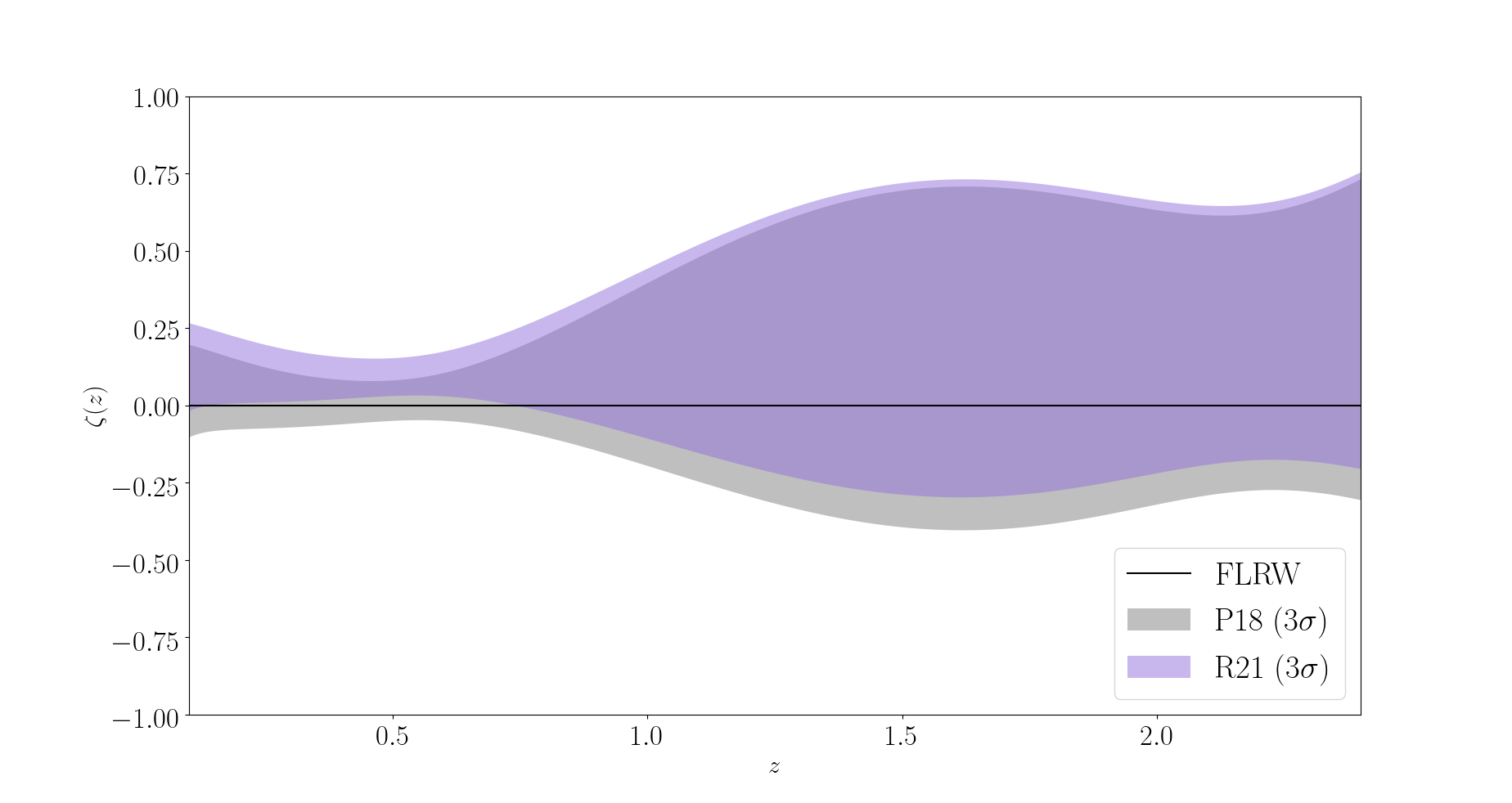}
\caption{Same as Fig.~\ref{fig:rec_null_test_kernel1}, but assuming the Mat\'ern(7/2) kernel  instead.}
\label{fig:rec_null_test_kernel2}
\end{figure*}

We also test how the null test results change with respect to different GP kernels. For instance, the reconstructions performed assuming the Mat\'ern(9/2) kernel shown in both panels of Fig.~\ref{fig:rec_null_test_kernel1} are
slightly more degraded compared to those obtained by the default kernel, which yielded in a decrease of the $\approx 3\sigma$ FLRW departure found previously for the C20-R21 and VBHJ17-P18 priors combination. When we assume the Mat\'ern(7/2) kernel, a similar trend is observed (see both panels of Fig.~\ref{fig:rec_null_test_kernel2}), as the reconstruction is a bit more degraded than the Mat\'ern(9/2) case, thus resulting in a further reduction of the null hypothesis deviation. Moreover, we caution that there are only a few data-points at lower redshifts - 3 transverse BAO measurements and 2 radial BAO ones - and hence we cannot ascribe statistical significance to any potential FLRW breakdown herein presented. We also find that these results are robust with respect to smaller number of reconstruction bins, e.g. 100 and 500 instead of 1000, and the inclusion of $H(z)$ measurements from galaxy ages. Although these data points do not come from BAO, they were added in order to check if the $H(z)$ measurements from radial BAO could bias our analysis towards a FLRW Universe, as they are obtained under the $\Lambda$CDM assumption - contrariwise to galaxy ages.

\section{Conclusions}

The Cosmological Principle, namely the assumption of large-scale homogeneity and isotropy of the Universe, comprises a fundamental assumption of modern cosmology. Such a hypothesis allows us to describe cosmological measurements by means of the FLRW metric. Nevertheless, it remains to be directly tested with cosmological observations in an extensive fashion. Because the caveats of the standard cosmological model remain unraveled, it is crucial to test whether they may arise due to an over-simplifying formulation of its foundations. 

Probing cosmic homogeneity is difficult due to the intrinsic limitation that we can only observe down the past light-cone. However, we can perform null tests such as $\zeta$. The radial and transverse mode of baryonic acoustic oscillations measurements must be consistent across the expansion history of the Universe, unless we have a violation of the FLRW assumption. In order to avoid further biases, we carry out a model-independent reconstruction of these data points across the redshift range using the Gaussian Processes. We assumed the best-fitted sound horizon scale from the transverse BAO mode to convert the angular scale into angular diameter distance measurements, in addition to different $H_0$ priors. 

Our analysis is in good agreement with the FLRW assumption at a $3\sigma$ confidence level at high redshifts. Mild deviations from this hypothesis are observed at lower redshifts ($0.1<z<0.3$) assuming the SH0ES $H_0$ prior, but not for Planck's, when we assume the $r_{\rm s}$ prior reported by~\cite{carvalho20}. This departure is slightly more pronounced assuming the sound horizon scale measurement from~\cite{verde17}, as the Planck's and SH0ES $H_0$ priors provided $\zeta(z) > 0$ at $0.1<z<0.3$ and $0.1<z<0.8$, respectively. We ascribe this signal to the limited data sampling herein used, since different choices Gaussian Processes kernels provided slightly more degraded reconstructions which reduced the statistical significance of these deviations. A thorough assessment on the priors impact in similar tests will be pursued in a future work, along with cosmological tests using these data-sets, and forecasts of such FLRW null test in light of future redshift surveys. We also found similar trends under the choice of different reconstructing binning scheme, and the inclusion of $H(z)$ measurements from galaxy ages. 

Therefore, our results help underpinning the Cosmological Principle as a valid physical assumption to describe the large-scale Universe, given the limitation of currently available observational data.  

\emph{Acknowledgments} -- 
CB is thankful to Rodrigo Von Marttens, Javier Gonzalez and Jailson Alcaniz for fruitful discussions. CB acknowledges financial support from the Programa de Capacita\c{c}\~ao Institucional PCI/ON, and from Funda\c c\~ao Carlos Chagas Filho de Amparo à Pesquisa do Estado do Rio de Janeiro (FAPERJ - fellowship {\it Nota 10}) at the late stage of this work. Code is provided at \url{https://github.com/astrobengaly/null_test_FLRW}


\end{document}